\documentclass[12pt]{article}
\usepackage{epsfig}
\usepackage{amssymb,graphicx}
\def\be{\begin{equation}}
\def\ee{\end{equation}}
\def\ba{\begin{array}{c}}
\def\ea{\end{array}}

\def\ben{\[}
\def\een{\]}

\newcommand{\bea}{\begin{eqnarray}}
\newcommand{\eea}{\end{eqnarray}}

\begin{document}

\begin{center}

{\Large \bf {

 Arnold's potentials and quantum catastrophes

 }}

\vspace{13mm}

\vspace{3mm}

\begin{center}

\textbf{Miloslav Znojil}\footnote{znojil@ujf.cas.cz}


\vspace{1mm}

The Czech Academy of Sciences, Nuclear Physics Institute,
 Hlavn\'{\i} 130,
250 68 \v{R}e\v{z}, Czech Republic

\end{center}

\vspace{3mm}

\end{center}

\subsection*{Keywords:}

Schr\"{o}dinger equation; multi-barrier polynomial potentials;
avoided energy-level crossings; abrupt wavefunction
re-localizations; quantum theory of catastrophes;

\subsection*{PACS number:}

PACS 03.65.Ge - Solutions of wave equations: bound states

\subsection*{Abstract}

In the Thom's approach to the classification of instabilities in
one-dimensional classical systems every equilibrium is assigned a
local minimum in one of the Arnold's benchmark potentials
$V_{(k)}(x)= x^{k+1} + c_1x^{k-1} + \ldots$. We claim that in
quantum theory, due to the tunneling, the genuine catastrophes (in
fact, abrupt ``relocalizations'' caused by a minor change of
parameters) can occur when the number $N$ of the sufficiently high
barriers in the Arnold's potential becomes larger than one. A
systematic classification of the catastrophes is then offered using
the variable mass term $\hbar^2/(2\mu)$, odd exponents $k=2N+1$ and
symmetry assumption $V_{(k)}(x)=V_{(k)}(-x)$. The goal is achieved
via a symbolic-manipulation-based explicit reparametrization of the
couplings $c_j$. At the not too large $N$, a surprisingly
user-friendly recipe for a systematic determination of parameters of
the catastrophes is obtained and discussed.

\newpage

\section{Introduction}

The infinite family of $(k-1)-$parametric polynomial potentials
 \be
 V^{(Arnold)}_{(k)}(x) = x^{k+1}+c_1\,x^{k-1} + c_2\,x^{k-2}
 + \ldots + c_{k-1}\,x
 \label{arno}
 \ee
was introduced, by Arnold \cite{Arnold}, as a natural generalization
of the four specific (sometimes called ``Lyapunov'') benchmark
functions which formed a background of the Thom's theory of
catastrophes \cite{Zeeman}. The theory was successful, in various
applications, mainly due to its identification of equilibria of
non-quantum, classical systems with coordinates  $x^{\rm
(equilibrium)}$ of the eligible minima of the
potentials~\cite{Thom}. Thus, the sudden losses of stability were
simulated and classified using, in Eq.~(\ref{arno}), the exponents
$k=2$ (the catastrophe called ``fold''), $k=3$ (the ``cusp''), $k=4$
(the ``swallowtail''), and $k=5$ (the ``butterfly''
catastrophe)~\cite{webcat}. In all of these models, the emergence
of a ``catastrophic'' instability was interpreted as a manifestation
of an abrupt disappearance of the relevant stable equilibrium, i.e.,
as a loss or merger of the real minima in the potential.

In the classical theory of catastrophes the losses or bifurcations
of equilibria were found caused by very small changes of parameters
$\{c_j\}$ in potential (\ref{arno}). It appeared tempting to try to
transfer such a concept of instability to quantum world. In one of
the recent implementations of such an idea~\cite{catast} the natural
analogue of the mergers of several local minima $V(x^{\rm
(equilibrium)}_j)$ has been sought in the non-approximative, {\em
exact\,} mergers of eigenvalues of a given observable at the so
called exceptional-point parameters~\cite{Kato} which are, in the
conventional quantum theory, not real~(cf. also, e.g., \cite{EP} for
further references).

In the latter approach (see also its sample implementations in
\cite{Ruzicka}) one had to resolve a number of rather difficult
conceptual as well as purely technical problems connected with the
mergers. Among them, a decisive obstacle lied in the difficulty of a
consistent probabilistic interpretation of the states at the complex
couplings. In the language of mathematics, it appeared necessary to
construct a fairly nontrivial physical Hilbert space (recommended
reading is the recent book \cite{book}). In applications, it proved
equally complicated to connect the schematic, mathematically
tractable benchmark models with the physical reality and experiments
\cite{Geyer}. Typically, the dynamical regime in which the
eigenvalues merged was produced using interactions which were, in
terms of observable coordinates, strongly non-local (detailed
explanation may be found, e.g., in review \cite{ali}).

For all of these reasons we decided to return to the roots by
keeping the couplings real. We shall merely require that the
imaginary parts of the (necessarily, complex) exceptional-point
couplings remain, for practical purposes, negligible. The key
inspiration of such a decision was provided by the elementary {\em
local\,} Thom's theory and by the Arnold's classification of the
classical catastrophes based on the use of Dynkin's diagrams and of
the related potentials (\ref{arno}) \cite{Arnold}.

In what follows we will start from the traditional concepts but we
will quickly move to their quantum-theoretical upgrades: Our systems
will be described by the bound-state Schr\"{o}dinger equation
 \be
 \left[- \frac{\hbar^2}{2\mu}\,\frac{d^2}{dx^2} +
 V^{(Arnold)}_{(k)}(x)
 \right ] \psi_n (x) =E_n\,
 \psi_n (x)
  \,, \ \ \ \ \
 n=0, 1, \ldots\,.
 \label{tiseh}
 \ee
As long as the interaction remains real, we are circumventing the
above-mentioned technical Hilbert-space difficulties: We will keep
working with wave functions $\psi_n (x)$ belonging to the
conventional physical Hilbert space $L^2(\mathbb{R})$. One should
only add that in the latter space the choice of potentials
$V^{(Arnold)}_{(k)}(x)$ with even $k=2N$ is not allowed because it
would leave the Hamiltonian unbounded from below. The consistent
Thom-inspired simulation of quantum catastrophes can only be based
on the asymptotically even potentials with $k=2N+1$ where $N = 1, 2,
\ldots$.

As a price to pay for our assumption of keeping our physical Hilbert
space $L^2(\mathbb{R})$ simple, the instant of the catastrophe will
be treated here as ``smeared''. In principle, naturally, one could
always try to amend the theory by its analytic continuation, in a
{\it Gedankenexperiment\,} at least (see, e.g., Refs.~\cite{ptho}
for a few elementary examples fulfilling the above-cited ambitious
requirement of having the crossings of the energy eigenvalues {\em
unavoided\,}). Nevertheless,  in this direction one could encounter
obstacles, not all of which seem to be resolved at present (see,
e.g., their incomplete list in \cite{MZbook}).

For all of these reasons we believe that for the majority of
practical purposes it makes good sense to insist on the reality,
locality and polynomiality of the benchmark-model interaction
potentials (\ref{arno}) in Schr\"{o}dinger equation (\ref{tiseh}).
On this background one can rely upon the conventional
density-distribution interpretation of the wave functions. In this
sense (cf. also a few more remarks in the first part of section
\ref{k2} below) we shall be interested in the observability of the
phenomena of the abrupt changes of the measurable features caused by
a small change of parameters $\{c_j\}$.

In the rest of section \ref{k2} we shall collect several methodical
comments and illustrate our present constructive
benchmark-model-analysis strategy on the most elementary double-well
special case of potential (\ref{arno}) with $k=2N+1=3$. In this
context we will emphasize that certain abstract mathematical
inconsistency of our present approximative, real-coupling ``avoided
crossing'' treatment of quantum catastrophes may be considered more
than compensated by its constructive and (not quite expected)
practical and strongly user-friendly features.

The first genuine illustration of the latter merits will be
presented, in section \ref{k3}, at $N=2$, i.e., using the
triple-well sextic-polynomial potential $V_{(5)}(x)= x^{6} +
c_1x^{4} + \ldots$. Among the specific merits of this ``quantum
butterfly'' model we will point out the extreme compactness of the
formulae (facilitating its purely numerical tractability in any
dynamical regime) as well as its methodical appeal: The model
provides, e.g., a very intuitive insight in the role of the size and
variability of the mass term ${\Lambda}^2=\hbar^2/(2\mu)$. One
should note that the choice of the units is already restricted here
by the fixed dominant-order coupling at $x^{k+1}$ in (\ref{arno}).

In subsequent section \ref{iuk4} the previous assertions will be
reconfirmed at $N=3$ (i.e., in essence, using $V_{(7)}(x)= x^{8} +
c_1x^{6} + \ldots$ in the quadruple-well dynamical regime). An
extension of some of these results beyond $N=3$ has been relocated
to Appendix A where some of the basic formulae are displayed up to
$N=8$, i.e., up to the potentials $V_{(17)}(x)= x^{18} + c_1x^{16} +
\ldots$.

In section \ref{k4} we will turn full attention to a few genuine
manifestations of the quantum tunneling and to the explicit
evaluation of its consequences, mainly in the dynamical regime of
deep wells separated by thick barriers. This will enable us to
complete the project and to describe, in the first nontrivial though
still non-numerical approximation, the phenomenon of a ``butterfly
relocalization'' quantum catastrophe at $N=2$. Finally, the
next-model $N=3$ parallels of this result will be described in
section~\ref{ck5}.

Multiple overall comments and summary will be added in
section~\ref{k6}.

\section{Preliminaries \label{k2}}

\subsection{Probability densities and their observability}

The most characteristic difference between the classical Thom's
catastrophe theory (which is, basically, a study of minima and
singularity theory {\it alias\,} an applied geometry of smooth
surfaces) and its tentative quantum-theoretical implementation lies
in the difference in the underlying concept of the measurement.
Indeed, in quantum systems in a stationary setup we, typically,
measure the energy levels as functions of the parameters. Naturally,
in such a setting we can hardly observe anything like a catastrophic
change of the scenario because in the space of the parameters the
energy levels form the curves without any special singularities.
Experimentally, even the apparent energy-level crossing phenomenon
is usually disproved via the mere enhancement of the precision of
the measurement.

The mathematical explanation of the paradox of the apparent
``repulsion'' of the levels is provided by the theory of the so
called exceptional points \cite{Kato}. It merely confirms that due
to the requirements of the unitarity of the evolution (i.e., of the
necessary self-adjointness of its generator called Hamiltonian) the
energy levels could only degenerate in the presence of a symmetry
\cite{Messiah}.

The chances of a feasible detection of a genuine quantum catastrophe
are much higher when one turns attention to the wave functions.
Thus, one should calculate and measure the wave-function-related
density of probability $\varrho(x)=\psi^*(x)\psi(x)$ of the
occurrence and of the spatial localization of particles in the
domains controlled by Schr\"{o}dinger equation.

\subsection{Spatially symmetric double well at $k=3$\label{upk2}}

In the Thom's classical theory of catastrophes the choice of the
$k=3$ special case of potential (\ref{arno}) [i.e., of the
two-parametric Lyapunov function $V^{(Arnold)}_{(3)}(x,c_1,c_2)$]
enables one to describe a bifurcation phenomenon {\it alias\,} one
of the classical catastrophes called cusp \cite{Thom}. The key
descriptive feature of potential $V^{(Arnold)}_{(3)}(x,c_1,c_2)$ is
that at negative $c_1=-a^2$ there emerges a central barrier which is
classically impenetrable. The system has a choice between the two
alternative equilibria.

After quantization, the tunneling is known to smear out the
possibility of such a bifurcation. In a more detailed explanation of
an apparent paradox let us consider Schr\"{o}dinger
Eq.~(\ref{tiseh}) with the potential in its spatially symmetrized
$c_2=0$ version. A phenomenologically most interesting double-well
shape is encountered,
 \be
 V(x)=V^{(Arnold)}_{(3)}(x,-a^{2},0)
 ={x}^{4}-a^{2}{x}^{2}
 \,.
  \label{[2]}
 \ee
The potential has a fixed local maximum in the origin ($V(0)=0$) so
that its double-well feature becomes more pronounced at the larger
$a \gg 1$. One then observes an approximate double degeneracy of the
low-lying spectrum. On both sides of the
barrier, at $x_{\pm}=\pm a/\sqrt{2}$, our potential possesses the
two equally deep negative minima $V(x_{\pm})=-a^4/4$. The low-lying
spectrum of bound states will, therefore, form the pairs $(E_0,
E_1)$, $(E_2, E_3)$, $\ldots$ with opposite spatial parities.
At the sufficiently large values
of $a \gg 1$
these
pairs will become almost, but never entirely, degenerate.

A slightly cumbersome occurrence of the numerical fractional factors
in the latter formulae can be suppressed. We get rid of these
redundant factors by the mere innocent-looking re-scaling
 \be
 a^2 = 2\,\alpha^2
 \label{innoc}
 \ee
of the coupling. The two equally deep local minima $V(\pm
\alpha)=-\alpha^4$ of the amended potential then lie at $x_{\pm}=\pm
\alpha$. Marginally, let us add that an analogous trick will be also
used at the higher exponents $k+1$ in Eq.~(\ref{arno}). Naturally,
the shortening and simplification of the formulae will be then
substantial.

\subsection{Deep-well approximation}

Whenever the parameters $a$ or $\alpha$ are sufficiently
large, the shape of the potential becomes well approximated, near
both of its minima, by the conventional Taylor expansions. In the
present case these expansions terminate yielding the exact
expressions
 \be
 V(\pm {\alpha}+y)=-{{\alpha}}^{4}
 +4\,{{\alpha}}^{2}{y}^{2}\pm 4\,{\alpha}{y}^{3}+{y}^{4}
 \,.
 \label{shift}
 \ee
In the units such that ${{\Lambda}}^2=\hbar^2/2\mu=1$ the
conventional Schr\"{o}{dinger equation for low-lying bound states
then splits, due to the thickness of the central barrier, into the
two approximately isospectral subproblems,
 \be
 \left[- \frac{d^2}{dy^2} -{{\alpha}}^{4}
 +4\,{{\alpha}}^{2}{y}^{2}\pm 4\,{\alpha}{y}^{3}+{y}^{4}
 \right ]\, \psi_n^{(\pm)} (y) =E_n^\pm\,
 \psi_n^{(\pm)} (y)
 \,, \ \ \ \ \
 n=0, 1, \ldots, n_{max}\,.
 \label{tise}
 \ee
After a rescaling $\ y \to z =\sqrt{2{\alpha}}\, y$, these equations
acquire the respective anharmonic-oscillator forms
 \be
 \left[- \frac{d^2}{dz^2}
 +{z}^{2}\pm \frac{1}{\sqrt{2{\alpha}^3}}\,{z}^{3}+\frac{1}{8{\alpha}^3}{z}^{4}
 \right ]\, \chi_n^{(\pm)} (z) =\frac{1}{2{\alpha}}(E_n^\pm+{\alpha}^4)\,
 \chi_n^{(\pm)} (z)\,.
 \label{tiseb}
 \ee
At the large ${\alpha}^2 \gg 1$ the leading-order formula is
obtained,
 \be
 E_{n}^\pm
  \approx -{\alpha}^4+2 (2n+1){\alpha}
  + {\cal O} (1/\sqrt{{\alpha}})\,,\ \ \ \ \
  n=0,1,\ldots, n_{max}\,
  \label{tisemaj}
 \ee
representing an approximate version of the two almost degenerate
low-lying eigenvalues $E_n^+ \lessapprox E_n^-$.

Needless to add, the precision of the latter asymptotic estimate may
systematically be improved, whenever needed, using the standard
formalism of Rayleigh-Schr\"{o}dinger perturbation theory in
application to the exact anharmonic-oscillator bound-state problem
(\ref{tiseb}).

\subsection{A note on the mass-dependence of the spectrum}

Before we move to the more general Schr\"{o}dinger equations with
$k>3$ in potential (\ref{arno}) let us note that our specific choice
of the trivial coupling at the asymptotically dominant term
$x^{k+1}$ does not imply any loss of generality. In place of the
changes of the coupling we will prefer a variation of the mass
$\mu$, i.e., of parameter ${\Lambda}^2={\hbar^2}/({2\mu})$. Firstly,
a genuine phenomenological appeal of such a convention can be seen
in the manifest clarity on the relationship between the large-mass
and semi-classical terminology as well as between the small masses
and certain ultra-quantum regime. Secondly, such a less standard
convention also proves useful from the point of view of mathematics.
Indeed, the introduction of the formally variable mass-term
coefficient ${\Lambda}^2$ in our ordinary differential
Schr\"{o}dinger equation enables us to replace formula
(\ref{tisemaj}) by its elementary generalization
 $$
 E_{n}^\pm \
  \approx -{\alpha}^4+2 {\Lambda} \,(2n+1){\alpha} + \ldots\,.
 $$
We immediately see that in the semiclassical limit the whole
low-lying spectrum converges to the minimum of the potential. The
smallness of ${\Lambda}$ implies a semiclassical behavior of the
spectrum, $\lim_{{\Lambda} \to
0}\,E_n({\Lambda})=\min\,V^{(Arnold)}_{(k)}(x)$. This provides a
direct contact with the classical catastrophe theory. One can also
notice that the quantum effects become more important in the
opposite extreme of large ${\Lambda}$, i.e., at the small masses
$\mu$.

In what follows, besides the conventional choice of ${\Lambda}=1$,
we intend to add also a systematic analysis of both of the latter
extremes. We feel motivated, first of all, by the fact that the
possible ``relocalization'' involving the quantum ground state might
be experimentally detectable. In the formal language the effect
would be caused by a small change of the parameters. Thus, there
willmanner be all reasons for speaking about the genuine quantum,
non-classical realizations of the intuitively appealing concept of a
catastrophe.

\section{$N-$plets of barriers and the localization at $N=2$\label{k3}}

In the Thom's list of one-dimensional Lyapunov functions
(\ref{arno}) one finds just the items with $k\leq 5$. This means
that in the classical theory of dynamical systems the $k= 5$ choice
called butterfly describes the last, most complicated scenario of
practical relevance. We have just shown that in quantum mechanics,
on the contrary, the solutions of Schr\"{o}dinger equation
(\ref{tiseh}) with small $k\leq 4$ are far from interesting. Hence,
it is fortunate that already at $k=5$ on can reveal the existence of
several nontrivial physical effects, some of them even in certain
thoroughly simplified special cases.

\subsection{Triple-well regime \label{bvpk3}}

The Arnold's general four-parametric function
 \be
 V^{(butter\!fly)}(x_1,c_1,c_2,c_3,c_4)
 = x^6 +c_1x^4 +c_2x^3  +c_3x^2 +c_4x\,.
 \label{butterb1}
 \ee
may describe a single-well, two-well or three-well potential in
general. Moreover, the vicinities of the minima can be characterized
by a wide variability of the parameter-dependent widths and depths
of the valleys. For our present purposes it will be sufficient to
study just the triple-well option, with the spatially symmetric
$V(x)=V(-x)$. Our preferred $k=5$ two-parametric, symmetric and
triple-well-admitting butterfly-related Lyapunov polynomial is
  \be
 V(x)=V^{(butter\!fly)}_{(5)}(x,-3a^2,0,3b^2,0)=x^6-3a^2
 x^4+3b^2x^2\,.
  \label{[6]}
 \ee
The purpose of the use of the specific augmented couplings is
twofold. Firstly, it extrapolates the recommended $k=3$
reparametrization~(\ref{innoc}) (simplifying some algebraic
manipulations) to the next, higher exponent $k$. Secondly, in a
parallel to the considerations of paragraph \ref{upk2}, it will also
help us to shorten the relevant formulae (typically, by eliminating
most of the redundant numerical coefficients).



\begin{figure}[h]                    
\begin{center}                         
\epsfig{file=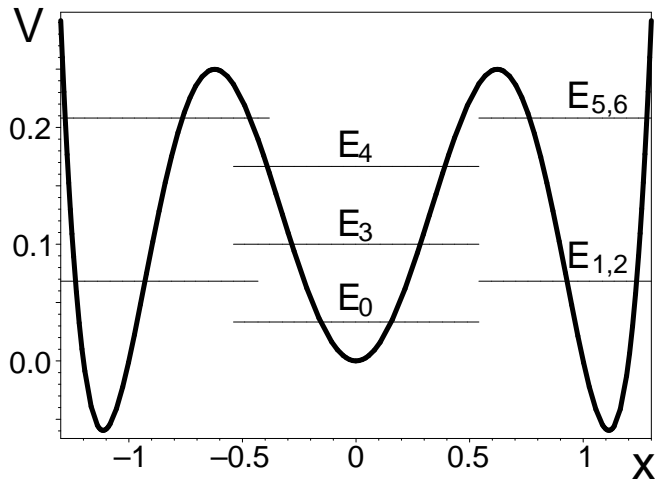,angle=270,width=0.56\textwidth}
\end{center}    
\vspace{2mm} \caption{Triple-well 
potential $ V(x)= x^6-61/25\,x^4+36/25\,x^2\,$ in units $\hbar^2=1$.
The lowest bound-state energy levels $E_n$  were calculated,
numerically, for a heavy particle of mass $\mu=18$.
 \label{triwja}
 }
\end{figure}

In the manner illustrated by Fig.~\ref{triwja} our present $k=5$
potential (\ref{[6]}) possesses, typically, the central local
minimum and the two pairs of the noncentral maxima and minima. Their
localization remains explicit and non-numerical because the
positions of these extremes coincide with the roots of polynomial
 \be
 V'(x)=6 x\, \left (
 {x}^{4}-2\,{a}^{2}{x}^{2}+{b}^{2}\right )\,.
 \ee
The two inner barriers have their maxima at
 \be
 x^{\rm (max)}_\pm=
 \pm \sqrt {{{a}^{2}-\sqrt
 {{a}^{4}-{b}^{2}}}}\,.
 \label{[9]}
 \ee
The two more relevant noncentral minima lie at equally compactly
defined
 \be
 x^{\rm (min)}_\pm=
 \pm \sqrt {{{a}^{2}+\sqrt
 {{a}^{4}-{b}^{2}}}}\,.
 \label{[10]}
 \ee
Already the simplicity of the latter formulae endorses fully the
introduction of the auxiliary numerical factors ``3'' in
Eq.~(\ref{[6]}). In addition, let us notice that all the four
extremes do exist (i.e., are real) whenever
 \be
 a^4>b^2\,.
 \label{sucha}
 \ee
In other words, their coordinates remain real whenever the
quartic-component coupling $a$ is chosen sufficiently large.

\subsection{Amended parametrization}

The specification of the parametric domain of interest (\ref{sucha})
defines a boundary curve which is rather cumbersome. The shape can
be simplified via another redefinition of the coupling constants.
After an introduction of a new {\it ad hoc\,} rescaling factor ``2''
(which will again simplify subsequent formulae and discussion) we
abbreviate
 \be
 [x^{\rm
 (max)}_\pm]^2=\alpha^2\,,\ \ \ \ \ [x^{\rm (min)}_\pm]^2=
 \alpha^2+2\,\beta^2\,.
 \label{fakk}
 \ee
This will certainly simplify the required guarantee of the
triple-well shape of the potential. Indeed, just the two
straight-line boundaries with $\alpha=0$ or $\beta=0$ will have to
be eliminated. The latter ansatz leads to our ultimate
parametrization of the potential, based on the inversion of the
mapping (\ref{[9]}) + (\ref{[10]}). What results are the
unexpectedly elementary formulae
 \be
 a^2=\alpha^2+\beta^2\,,
 \ \ \ \
 b^2={\alpha^2}\,
 (\alpha^2+2\,\beta^2)\,.
 \ee
Their use keeps the height of the barriers elementary,  growing with
both $\alpha^2$ and $\beta^2$. At the two maxima we have
 \be
 {\rm }\ \  \ V(\pm \alpha)={\alpha}^{4}(\alpha^2+3\,{\beta}^{2})\ >\ 0\,.
 \ee
The parallel explicit expression for the depths of the potential at
its off-central minima $
 x^{\rm (min)}_\pm=\pm
R$ with $R= \sqrt{\alpha^2+2 \,\beta^2}$ appears equally compact and
user-friendly. At the two off-central minima we have
 \be
 {\rm }\ \  \
 V(\pm R )
 ={\alpha}^{6}+3\,{\alpha}^{4}{\beta}^{2}-4\,{\beta}^{6}
 =\left (\alpha^2-\beta^2 \right)\,R^4\,.
 \ee
This value is positive when $\alpha^2 > \beta^2 $ (the extremes are
then higher than the central absolute minimum
with $V(0)=0$), or negative when $\beta^2>\alpha^2 $ (then, we deal
with the two equal absolute minima of the potential).

\subsection{Numerical spectra and localization}


A decisive encouragement of our present non-numerical constructive
project emerged during a routine, purely numerical analysis of
quantum bound states supported by the Thom's most complicated
four-parametric potential (\ref{arno}) with $k=5$. The problem under
consideration was the role of the mass-term coefficient
${{\Lambda}}^2=\hbar^2/(2\mu)$. In one extreme (viz., in the
semiclassical regime with small ${\Lambda}$) the results (viz., the
decrease of the levels to the minimal of the valleys) were
expectable. No surprises were encountered in the domain of
${\Lambda} \approx 1$ (reason: this value is used, in many
computations, after the most popular choice of the units such that,
strictly, ${\Lambda}=1$). We were, therefore, interested in the
light-particle-motion regime in which the quantum effects (like,
e.g., tunneling) get enhanced due to the related increase of the
distance between the individual energy levels.

A characteristic sample of our results is given here in
Fig.~\ref{triwja} where the shape of the potential with couplings $
c_1=-61/25 $, $ c_2=0 $, $ c_3=36/25 $ and $ c_4=0$ is displayed.
After we choose a small mass such that
$1/{{\Lambda}}^2=2\mu/\hbar^2=36$, a characteristic low-lying
``quantum butterfly'' bound-state spectrum was obtained which
already lied high over the minima of the respective valleys in the
potential. In the picture we see the seven lowest numerically
evaluated energy levels. The following interesting spectral features
of the model can be deduced.

 \begin{itemize}
 \item
 The ground state is localized in the central well which
 is not the deepest one.
 This is an apparent paradox which finds its explanation in the
 fact that the other two deeper wells are, in comparison, much narrower.

 \item
 The first two excited states with energies $E_1$ and $E_2$
 (as well as the two higher states with $E_5$ and $E_6$)
 are almost degenerate (in the picture we do not see their
 exponentially small difference).
 They are also predominantly localized out of the central well.
The observation sounds like a paradox because the minima are the
absolute minima of the potential. The explanation of the puzzle is
easy because near these minima the potential well itself is very
narrow and steep.

 \item
A deeper scrutiny of the situation reveals, in addition, another
apparent conflict with the conventional wisdom because the lower,
single-node state $\psi_{1}(x)$ is spatially antisymmetric while its
slightly higher two-nodal second-excitation partner $\psi_{2}(x)$ is
spatially symmetric.
 This remains compatible with the
 observation that the
 (single) nodal zero
 of the spatially antisymmetric wavefunction $\psi_1(x)$ is
 ``remote'' (i.e.,
 that it lies in the origin).

 \end{itemize}

 \noindent
Up to the outer-wells quasi-degeneracies our choice of the
parameters keeps the individual energy levels sufficiently separated
when assigned to one of the spatial (i.e., inner or outer)
arrangements. At the same time, there is no guarantee of a
separation between the inner and outer levels. This observation
points at our present key idea: One can certainly change the set of
the coupling constants in such a way that the central minimum moves
upwards or, alternatively or simultaneously, the outer two minima
move down.

One must expect that at a certain critical value of the parameters
the energies $E_0$, $E_1$ and $E_2$ get very close to each other. At
such an apparent level-crossing instant, what should be expected is
a change of the ground state and of its central localization to the
two distinct, spatially separated outer-minima vicinities. The role
of the ground state then gets transferred from the wavefunction
$\psi_0$ of Fig.~\ref{triwja} to another, spatially symmetric
wavefunction which would be nodeless but localized in the outer two
wells. Due to the quasi-degeneracy, also the role of the first
excited state will be played by the (newly, antisymmetric) partner
of the ground state. The lowest state with the centrally localized
wavefunction will only represent the second excitation.

%
%
%

\section{Four-well model with $N=3$\label{iuk4}}

Out of the Arnold's general family (\ref{arno}) let us pick up the
potential with $k=7$. Its four-well shape can be then guaranteed
using its three-parametric even-parity octic-polynomial special case
 \be
 V(x) = x^8-4a^2x^6+6b^2x^4-4c^2x^2\,.
 \label{[20]}
 \ee
This is a simplified Lyapunov function (see its graphical sample
with $a=2$, $b=\sqrt{13}$ and $c=2\sqrt{7}$ in Fig.~\ref{fifi}), the
use of which will still be sufficient for our present purposes of a
fully quantitative analysis of the possible quantum-catastrophic
evolution scenarios.

%
%
%
\begin{figure}[h]                    
\begin{center}                         
\epsfig{file=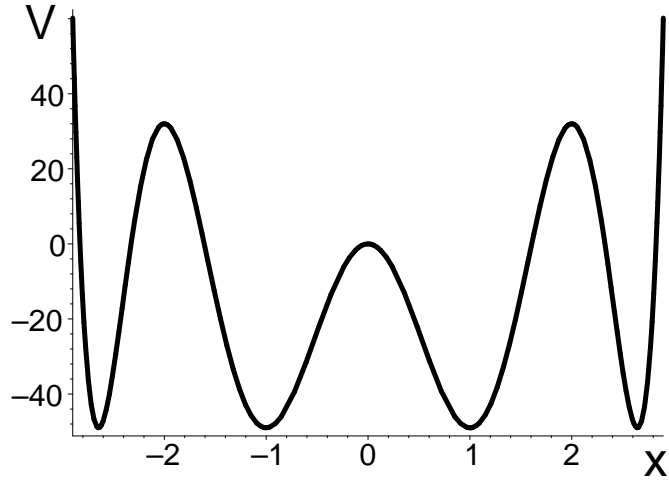,angle=270,width=0.56\textwidth}
\end{center}    
\vspace{2mm} \caption{Four-well potential (\ref{[20]}) with  $N=3$.
 \label{fifi}
 }
\end{figure}


%
%
%
%
%
%
%
%
%
%
%
%

\subsection{Parameters}

Besides the trivially localized central maximum $V(0)=0$ we need to
localize the remaining six extremes of this function. This means
that we need to find the zeros $x^{(max/min)}_\pm = \pm
\sqrt{\xi^{(max/min)}}$ of its derivative $V'(x)$, i.e., the three
positive roots of the following cubic polynomial
 \be
  C^{(3)}(\xi)=\xi^3-3a^2\xi^2+3b^2\xi-c^2\,.
  \label{21}
 \ee
The direct-solution method used in the preceding section would be of
little help now because the well known Cardano formulae
are overcomplicated,
expressing the required
roots as differences between
auxiliary complex numbers.
Moreover, we found that also the {\em ad hoc\,}
positive-root recipe
of Ref.~\cite{ss11ss} is not suitable for our present purposes.
Fortunately,
what we found efficient was the idea of the
reparametrization of the couplings. Thus,
in a generalization of the recipe
of the preceding subsection we now introduce the
real triplet of parameters $(\alpha,\beta,\gamma)$ such that
 \be
  C^{(3)}(\xi)=(\xi-\alpha^2)(\xi-\alpha^2- 3\beta^2)
 (\xi-\alpha^2- 3\beta^2-3\gamma^2)
  \,.
  \label{elik}
 \ee
This product may be expanded,
 \ben
 C^{(3)}(\xi)={\xi}^{3}+ \left( -3\,{\alpha}^{2}-6\,{\beta}^{2}-3\,{\gamma}^{2}
 \right) {\xi}^{2}+ \left( 12\,{\alpha}^{2}{\beta}^{2}+6\,{\alpha}^{2}
{\gamma}^{2}+9\,{\beta}^{4}+3\,{\alpha}^{4}+9\,{\beta}^{2}{\gamma}^{2}
 \right) \xi-
 \een
 \be
 -9\,{\alpha}^{2}{\beta}^{2}{\gamma}^{2}-{\alpha}^{6}-6\,{
\alpha}^{4}{\beta}^{2}-3\,{\alpha}^{4}{\gamma}^{2}-9\,{\alpha}^{2}{
\beta}^{4}\,.
 \ee
The comparison with Eq.~(\ref{21}) immediately leads to the required
reparametrization formulae. For the triplet of the present couplings we get
 \ben
 a^2={\alpha}^{2}+2\,{\beta}^{2}+{\gamma}^{2}\,,
 \ \ \ \
 b^2={\alpha}^{4}+4\,{\alpha}^{2}{\beta}^{2}+2\,{\alpha}^{2}
{\gamma}^{2}+3\,{\beta}^{4}+3\,{\beta}^{2}{\gamma}^{2}\,,
 \een
 \be
 c^2={\alpha}^{2}\,
 \left ({\alpha}^{4}+6\,{
 \alpha}^{2}{\beta}^{2}+3\,{\alpha}^{2}{\gamma}^{2}
 +9\,{
 \beta}^{4}
 +9\,{\beta}^{2}{\gamma}^{2}
 \right )\,.
 \label{[27]}
 \ee
%
%
In particular, we have $\alpha=\beta=\gamma=1$ in Fig.~\ref{fifi}.

\subsection{Scaling\label{uhu}}

In Eqs. (\ref{fakk}) and (\ref{elik}) our choice of the redundant
numerical coefficients was intuitive. In retrospective, the
auxiliary rescaling may be validated by elementary combinatorial
analysis. The aim of such an analysis is to find an analogous
optimal parametrization ansatzs at the higher odd degrees $k>7$. At
$k=7$ itself the elimination of the fractional coefficients may be
sought via the following generalized ansatz (\ref{elik}),
 \be
  C^{(3)}(\xi)=(\xi-\alpha^2)(\xi-\alpha^2- P\beta^2)
 (\xi-\alpha^2- Q\beta^2-R\gamma^2)
  \,.
  \label{gelik}
 \ee
where $P$, $Q$ and $R$ have to be (small) integers. As a polynomial
in $\xi$ this expression reads
 \ben
 C^{(3)}(\xi)={\xi}^{3}+ \left( -{\it {P}}\,{\beta}^{2}
 -3\,{\alpha}^{2}-{\it {Q}}\,{
 \beta}^{2}-{\it {R}}\,{\gamma}^{2} \right) {\xi}^{2}+
 \een
 \ben
 +\left(
 2\,{\alpha}^{2} {\it {Q}}\,{\beta}^{2}+2\,{\alpha}^{2}{\it
 {R}}\,{\gamma}^{2}+3\,{\alpha}^{4} +{\it {P}}\,{\beta}^{4}{\it {Q}}+{\it
 {P}}\,{\beta}^{2}{\it {R}}\,{\gamma}^{2}+ 2\,{\alpha}^{2}{\it
 {P}}\,{\beta}^{2} \right) \xi- \ldots\,
 \een
having to match Eq.~(\ref{21}). Thus, the ${\cal O}(\xi^0)$
component was omitted as irrelevant while the coefficients at $\xi$
and $\xi^2$ have to be, in the light of Eq.~(\ref{21}), divisible by
three. Besides the obvious most elementary choice of $R=3$, this
implies that we have to choose $P$ and $Q$ such that both their sum
and product become divisible by three, i.e., $P+Q=3m$ and $P Q=3n$.
There are no such integers at $m=1$ while at $m=2$ the unique
solution of this diophantine problem is $P=Q=3$.

\subsection{Barriers and valleys.}

Given the optimal parametrization at $k=7$, the innermost pair of
minima occurs at the coordinates $x=x_{inn.\ min}=\pm \alpha$. They
are always negative:
 \be
 V(\pm \alpha) = - \left( {\alpha}^{4}
 +8\,{\alpha}^{2}{\beta}^{2}+4\,{
 \alpha}^{2}{{\gamma}}^{2}
 +18\,{\beta}^{4}+18\,{\beta}^{2}{{\gamma}}^{2}
  \right){\alpha}^{4}\,.
 \ee
The depth of these wells grows with all of the parameters, still
with the decisive role played by the growth of $\alpha$.

The subsequent intermediate maxima of $V(x)$ occur at $x=x_{loc.\
max}=\pm T$,
 \be
 V(\pm T)=
 \left[3\,{\beta}^{4} +6\,{\beta}^{2}{{\gamma}}^{2
 }-{\alpha}^{2}
 \left ({\alpha}^{2}
 +2\,{\beta}^{2}+
 4\,{{\gamma}}^{2}\right )
  \right]\,T^4\,.
 \ee
We abbreviated here $T=T(\alpha,\beta)=\sqrt
{{\alpha}^{2}+3\,{\beta}^{2}}$. At small $\alpha$ the height of
these barriers will be positive and growing with $\beta$. The sign
of their height determines their dominance or subdominance in
comparison with the central local maximum $V(0)=0$. This sign can be
predetermined by an {\it ad hoc\,} restriction on the size of
$\alpha$.

The remaining, outer pair of minima lies at $x=x_{out.\ min}=\pm R$,
with the values of potential
 \be
 V(\pm R) = -\left({\alpha}^{4}
 +2\,{\alpha}^{2}{\beta}^{2}
  +3\,{{\gamma}}^{4}
  -3\,{\beta}^{4} -2\,{\alpha}^{2}{{\gamma}}^{2}
 \right)  \,R^4
 \ee
where $ R=R(\alpha,\beta,\gamma)= \sqrt
{{\alpha}^{2}+3\,{\beta}^{2}+3\,{{\gamma}}^{2}}$. The sign as well
as the depth of the outer minima can be best controlled by the
variation of the magnitude of $\gamma$.

\section{Bound states \label{k4}}

The existence of the
``realistic'' wavy shapes of the Arnold's potentials $V(x)$
with $k=2N+1$ growing from $N=1$ (section \ref{k2})
to $N=2$ (section \ref{k3})
to $N=3$ (section \ref{iuk4})
is guaranteed by their present specific parametrization
in terms of the coordinates of the extremes.
Let us add that the resulting compact formulae
for the minima and maxima
make even the Arnold's $k=2N+1$ model with $N=3$ user-friendly.

In practice, naturally,
one has to solve the underlying Schr\"{o}dinger equation
by the suitable {\it ad hoc\,} numerical methods in general.
Then, the availability of the closed formulae
certainly simplifies the task.
For illustration,
a few purely numerical bound-state energies were
also sampled above, in Fig.~\ref{triwja}, at $N=2$.

From the point of view of applied quantum physics,
the growth of $N$ leading to a
more and more oscillatory shape of $V(x)$
might also prove important as
mimicking physical reality.
For example, one can have in mind an apparently
periodic physical system which is in fact far from infinite
(cf., e.g., small crystals or not too large
quasi-one-dimensional molecules \cite{PRB}, etc).
Naturally, the description of such a system would
require a transition from conventional periodic potentials
to some more realistic forces exhibiting just a
finite number of minima.
For this reason we constructed and described,
in Appendix A,
the five
further parametrizations of $V(x)$ at $N=4,5,6,7,$ and $8$.

Needless to add that the user-friendliness of some of these
models may still be surprising.
For
example, in a trial-and-error search for the equal-depth coincidence
$V(R) =V(\alpha)$ at $N=3$ we found the amazingly
elementary solution $\alpha=\beta=\gamma=1$.
We choose also these remarkable parameters in the sample of
the shape of $V(x)$ in Fig.~\ref{fifi} above.

\subsection{Single-barrier tunneling ($k=3$)}

For our present purposes of the mere qualitative characterization of
the phenomenon of quantum catastrophes we shall find it sufficient
to use the non-numerical, {\em approximate\,} means of perturbation
theory. Our first result of such a type was presented in section
\ref{k2}. In the double-well potential (\ref{[2]}) we redefined $x =
\pm \alpha + y$. This moved the origin of the axis of coordinates
from the ``useless'' center of spatial symmetry $x=0$ [with the
local maximum of our double-well potential] to the coordinate $y=0$
of one of the local minima. The shift redefined the couplings [see
Eq.~(\ref{shift})] and converted our initial, purely numerical bound
state problem (\ref{tise}), in the special case of the large
parameter $\alpha \gg 1$, into a user-friendlier, perturbatively
solvable Schr\"{o}dinger Eq.~(\ref{tiseb}).

As long as our Schr\"{o}dinger equation acquired a weakly perturbed
harmonic oscillator form possessing the well known leading-order
solutions (\ref{tisemaj}), we revealed an impossibility of any
bifurcation at $k=3$. The ``cusp'' catastrophe known from classical
physics disappeared after quantization. Due to the emergence of the
tunneling through the central barrier, the quantum state (of a given
spatial parity) has been found localized, {\em simultaneously}, near
{\em both\,} of the minima of the potential.

\subsection{Bifurcation ($k=5$)} 

In section \ref{k2} we emphasized that the Thom's classical
catastrophe called ``cusp'' does not possess a quantum analogue due
to tunneling. The simplest eligible candidate for a
``catastrophe-simulating'' {\it alias\,} bifurcation-admitting
quantum Arnold's potential must be sought and can be found at the
next exponent $k=2N+1=5$.

\subsubsection{Central-well-supported wave functions and energies.}

The leading-order formula for the anharmonic-oscillator central-well
part of the spectrum is well known,
 \be
 E_m^{(central)}= \sqrt{3}\, (2m+1)\,b + \ldots \,,
 \ \ \ \ m = 0, 1, \ldots\,,
 \ \ \ \ b = \alpha\,\sqrt{\alpha^2+2\,\beta^2}
 \,.
 \label{central5}
 \ee
We will assume that the value of $\beta$ is large but we also allow
the growth of the magnitude of $\alpha= \mu\,\beta$ with, say, $\mu
\leq 1$. The two parameters will coincide in the limit $\mu \to 1$
of course. Once we insert this ansatz in the couplings we get
$a=\beta\sqrt{\mu^2+1}$ and $b= \mu \beta^2\sqrt{\mu^2+2}$ in
(\ref{[6]}). With the position $R=\beta\,\sqrt{\mu^2+2}$ of the
off-central minimum of the potential we may expect the existence of
a critical value of $\mu=\mu^{(critical)}$ at which the ground-state
energy~(\ref{central5}) would cross the lowermost double-well
eigenvalue as given by Eq.~(\ref{groudw}) below.

\subsubsection{Double-well-supported wave functions and energies.}

In an immediate vicinity of the two non-central
quantum-butterfly-model minima (\ref{[10]}), local or absolute, we
may introduce a shifted coordinate $y$ and get the two new,
equivalent and exact representations of the potential,
 \ben
 V(\pm R +y)
 =
 V(\pm R )+ 12\,R^2\,\beta^{2} {y}^{2} \pm 4\,R\,
 \left( 2\, R^{2}+3\,{\beta}^{2} \right) {y}^{3}+
\left( 12\,R^{2}+ 3\,{\beta}^{2} \right) {y}^{4}\pm 6\,R\,
{y}^{5}+{y}^{6}\,.
 \een
The shift leads to the two alternative Schr\"{o}dinger equations
 \be
 \left[- \frac{d^2}{dy^2}
 + 12\,R^2\,\beta^{2} {y}^{2} \pm 4\,R\,
 \left( 2\, R^{2}+3\,{\beta}^{2} \right) {y}^{3}+
 \ldots
 \right ]\, \psi_n^{(\pm )} (y) =
 \left (E_n^{(\pm )} -V(\pm R)
 \right )\,
 \psi_n^{(\pm)} (y)
  \label{tisec}
 \ee
where $ n=0, 1, \ldots\,$ numbers the approximately degenerate
energy doublets.

A simplified picture of dynamics is obtained in the deep
double-well-dominated regime with $\beta^2>\alpha^2$, i.e.,
typically, at an arbitrary $\alpha^2$ and at a sufficiently large
$\beta^2\gg 1$. In this regime the minima of the potential are very
deep ($V(\pm R) = -4\beta^6 + {\cal O}(\beta^2)$) and very narrow
($R^2\beta^2 = {\cal O}(\beta^4)$). The rescaling $\ y \to z
=\varrho\, y$ with $\varrho^4=1/(12\,R^2\,\beta^{2})$ then converts
Eq.~(\ref{tisec}) into anharmonic-oscillator bound-state problem
 \be
 \left[- \frac{d^2}{dz^2}
 +{z}^{2}\pm \lambda\, {z}^{3} + \ldots
 \right ]\, \chi_n (y) =\varrho^2\,(E_n -V(\pm R))\,
 \chi_n (y)
 \,,
 \ \ \ \ \ \ \lambda=4\,R\,\varrho^5\,
 \left( 2\, R^{2}+3\,{\beta}^{2} \right)\,.
 \label{tisebd}
 \ee
As long as $2\beta^2 < R^2 < 3\beta^2$, the size of the perturbation
is asymptotically negligible,
 \be
 \frac{7\varrho }{3R}\ < \ \lambda\  < \ \frac{9\varrho }{3R}
 \ = \ {\cal O} (1/\beta^2 )\,.
 \ee
Thus, at $\beta^2 \gg \alpha^2$ the anharmonicity induces just a
small perturbation correction to the dominant, almost degenerate
even- and odd-parity energies $E_n^{(+)}\ \lessapprox \ E_n^{(-)}$,
 with
 \be
 \ E_n^{(\pm)} \approx
 V(R)+\sqrt{12}\,(2n+1)\, R \beta + {\cal O}(\beta^{-2})\,,
 \ \ \ \ n  = 0, 1,  \ldots, n_{max}\,.
 \label{groudw}
 \ee
The error estimate should, in principle, reflect the possible ${\cal
O}(\beta^{0})$ influence of the cubic anharmonicity. Nevertheless,
as long as function $y^3$ is spatially antisymmetric, this
correction vanishes in the first order approximation. For this
reason the error term vanishes asymptotically.

\subsection{Avoided crossings}

The comparison of formulae (\ref{groudw}) and (\ref{central5})
reveals that in the asymptotic region (i.e., in the limit $\beta \to
\infty$) the decisive role is played by the change of the sign of
the depth $V(R)= {\cal O}(\beta^6)$ of the outer minima of the
potential itself. Trivially we get
 \be
 \mu^{(critical)}=1\,.
 \label{theasy}
 \ee
A systematic inclusion of the higher-order corrections also remains
feasible. For example, up to the subdominant order of magnitude
${\cal O}(\beta^2)$ we have the closed pair of formulae concerning
the respective even-parity ground-state energies
 \be
 E_0^{(single-well)}=\sqrt{3}\alpha\,R\,,\ \ \ \ \ \
 E_0^{(double-well)}=
 (\alpha^2-\beta^2)\,R^4+2\sqrt{3}\beta\,R \,.
 \label{punta}
 \ee
An important physical motivation of such an enhancement of the
precision lies in the fact that the latter formula already offers
the predictions which differ from the classical picture. Still, we
see that in comparison with the classical signature of the
catastrophe (marked by the change of sign of the position of the
minimum $V(R)$), the quantum effect remains small.

We believe that it is still instructive to perform the analysis in
some detail because the underlying avoided-level-crossing phase
transition has a clear physical interpretation of a sudden change of
the localization of the wave functions between the {\em two\,}
spatially well separate vicinities of the off-central, {\em
remote\,} minima (with $x \approx \pm R$) at $\mu < 1$, and the {\em
single\,}, very small vicinity of the origin (i.e., to $x \approx
0$) at $\mu > 1$. Such an abrupt, ``quantum-catastrophic'' change of
the localization of the ``butterfly'' quantum system in question
would certainly be measurable. Naturally, for the non-asymptotic
parameters $\alpha$ and $\beta$, an exact and truly reliable
determination of the value of $\mu^{(critical)}$ remains purely
numerical.

Only some of the qualitative aspects of the parametric dependence of
the level-crossing instant can be offered by perturbation theory. An
important merit of its second-order form (\ref{punta}) should be
seen in its simplicity. Out of the difference
 \ben
 \Delta=E_0^{(double-well)} -E_0^{(single-well)}
 \een
one can factor out $\sqrt{3}\,R$ and obtain the catastrophic locus
as an implicitly, numerically defined curve $\Delta=0$ as displayed
in Fig.~\ref{spja}.

%
\begin{figure}[h]                    
\begin{center}                         
\epsfig{file=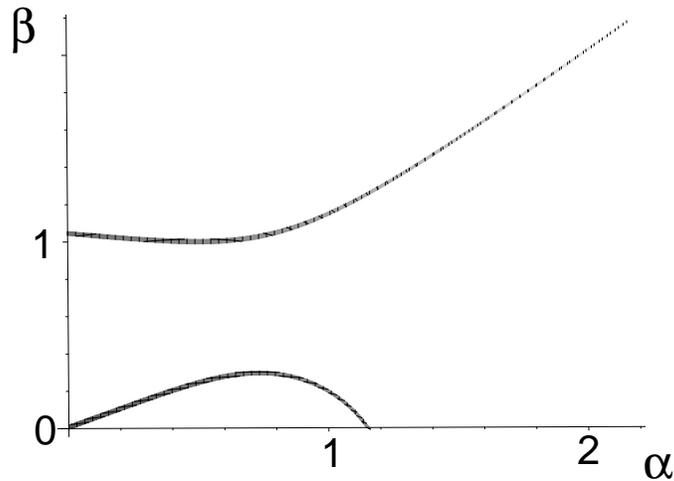,angle=270,width=0.56\textwidth}
\end{center}    
\vspace{2mm} \caption{The upper and lower bound $\beta=\beta^\pm
(\alpha)$ of the domain $D$ of dominance of the single-well,
centrally localized ``butterfly''  ground state over its double-well
alternative in zero-order approximation (\ref{punta}). Beyond
$\alpha \approx 1.2$ the deviations from the asymptotic
$\beta=\alpha$ [cf Eq.~(\ref{theasy})] are surprisingly small.
 \label{spja}
 }
\end{figure}

The main message provided by this picture is that the deviations
from the asymptotic phase-transition straight line
$\beta^{(catastrophic)}(\alpha)=\alpha$ do not seem to become too
large even in the deeply non-asymptotic domain of parameters
$\alpha$ and $\beta$. In addition, we found it rather surprising
that besides an expected, smooth continuation of the asymptote
$\beta^{(catastrophic)}(\alpha)$ there also exist certain
``anomalous'' quantum-catastrophe roots of the difference $\Delta$.
In Fig.~\ref{spja} they form a second, lower boundary. At the
$\beta$s below this curve the double-well ground state returns to
the dominance in spite of the positivity of the double-well minima.
In fact, the effect is not entirely surprising because it just
reflects the larger width of the two relevant non-central valleys.

\section{Avoided crossings at $k=7$\label{ck5}}

As long as Eq.~(\ref{punta}) only compares the two competing
candidates for the global ground state energy in the leading-order
perturbation approximation, the pattern displayed in Fig.~\ref{spja}
might easily change after the necessary step-by-step inclusion of
the higher-order corrections into the separate candidates for the
energies. At the non-asymptotic, small $\alpha$ and $\beta$, in
particular, the role of the anharmonicities in $V(x)$ will also
increase. Still, the main surprise accompanying the preceding
perturbation analysis considerations may be seen in the utterly
unexpected compactness and simplicity of the formulae. The merit
seems to survive the transition not only to the higher orders of
perturbation theory but also, in parallel, to the Arnold's
polynomials of higher degrees. For confirmation let us now pick up
$k+1=8$.

%





\subsection{Inner minima}

In terms of parameters (\ref{[27]}) the shape of our potential
(\ref{[20]}) exhibits the pair of the ``inner'' minima at $x=\pm
\alpha$, with the equal depths
 \be
 V(\pm \alpha)=-\left( {\alpha}^{4}+8\,{\alpha}^{2}{\beta}^{2}+4\,{
 \alpha}^{2}{\gamma}^{2}+18\,{\beta}^{4}+18\,{\beta}^{2}{\gamma}^{2} \right)\,
 \alpha^{4}\,.
 \label{35}
 \ee
In the vicinity of both of these minima we
may truncate the Taylor-series expansion and obtain the
potential in its
leading-order harmonic-oscillator approximation
 \be
 V(\pm \alpha+y)=V(\pm \alpha)+
  \omega^2\,y^2+{\cal O}(y^3)
  \,,
 \ \ \ \
 \omega^2=72\,{\alpha}^{2}{\beta}^{4}+72\,{\alpha}^{2}{\beta}^{2}{{\gamma}}^{2}
 =72\,\beta^2
 \left( {\beta}^{2}+{{\gamma}}^{2} \right)  \alpha^2
  \ee
Beyond this approximation we only need to go up to the eighth power
of $y$ in order to obtain the exact, zero-error formula
 \ben
 V(\pm \alpha+y)=
 {y}^{8}\pm 8\,\alpha\,{y}^{7}+ \left( -8\,{\beta}^{2}-4\,{g}^{2}+24\,{
 \alpha}^{2} \right) {y}^{6}\pm  \left(
 32\,{\alpha}^{3}-24\,{g}^{2}\alpha -48\,{\beta}^{2}\alpha \right)
 {y}^{5}+
 \een
 \ben
 + \left( -48\,{\alpha}^{2}{g}^{2
 }+16\,{\alpha}^{4}-96\,{\alpha}^{2}{\beta}^{2}+18\,{\beta}^{2}{g}^{2}+
 18\,{\beta}^{4} \right) {y}^{4}\pm
 \een
 \ben
 \pm \left(
 72\,{\beta}^{4}\alpha-32\,{g}^
 {2}{\alpha}^{3}-64\,{\beta}^{2}{\alpha}^{3}+72\,{\beta}^{2}{g}^{2}
 \alpha \right) {y}^{3}+
 \een
  \be
   +\left(
 72\,{\alpha}^{2}{\beta}^{4}+72\,{\alpha }^{2}{\beta}^{2}{g}^{2}
 \right) {y}^{2}-4\,{g}^{2}{\alpha}^{6}-18\,{
 \beta}^{4}{\alpha}^{4}-8\,{\beta}^{2}{\alpha}^{6}-18\,{\beta}^{2}{g}^{
 2}{\alpha}^{4}-{\alpha}^{8}\,.
 \ee



\subsection{Outer minima}



In our parametrization the ``outer'' pair of minima lies at $\pm R=
\pm \sqrt{\alpha^2+3\,\beta^3+3\,\gamma^2}$, with
 \be
 V(\pm R)=- \left( 3\,{\gamma}^{4}-2\,{\alpha}^{2}{\gamma}^{2}
 +2\,{\alpha}^{2}{\beta}^{2}-
 3\,{\beta}^{4}+{\alpha}^{4} \right)  R^{4} \,.
 \label{37}
 \ee
In the vicinity we have the leading-order harmonic-oscillator
approximation again,
 \ben
 V(\pm R+y)=V(\pm R)+
  \Omega^2\,y^2+{\cal O}(y^3)\,,
  \een
  \be
 \Omega^2=72\,{\alpha}^{2}{\beta}^{2}{{\gamma}}^{2}
 +216\,{{\gamma}}^{6}+216\,{\beta}^{4}{{\gamma}}^{2
 }+432\,{\beta}^{2}{{\gamma}}^{4}+72\,{\alpha}^{2}{{\gamma}}^{4}
 =72\,{{\gamma}}^{2} \left( {\beta}^{2}+{{\gamma}}^{2} \right)
 R^2\,.
 \ee
Again, the application of the higher-order Rayleigh-Schr\"{o}dinger
perturbation expansions may be based on the use of the following
exact, octic-anharmonic-oscillator potential
 $$
 V(\pm R+y)=
 {y}^{8}\pm 8\,{R}{y}^{7}+
 \left(76\,{\beta}^{2}+24\,{\alpha}^{2}+80\,{g}^{2} \right) {y}^{6}\pm
  \left( 144\,{g}^{2}+32\,{\alpha}^{2}+120\,{\beta}^{2} \right)\,R\,  {y}^{5}+
 $$
 $$
 +
 \left(
 192\,{\alpha}^{2}{g}^{2}+738\,{\beta}^{2}{g}^{2}+450\,{g}^{4}+144\,{
 \alpha}^{2}{\beta}^{2}+288\,{\beta}^{4}+16\,{\alpha}^{4}
 \right) {y}^{4}\pm
 $$
 $$
 \pm \left(360\,{\beta}^{2}{g}^{2}+64\,{\alpha}^{2}{g}^{2}+264\,{g}^{4}+96\,{
\beta}^{4}+32\,{\alpha}^{2}{\beta}^{2}
 \right)\,R\, {y}^{3}+
  \Omega^2\,y^2+V(\pm R)\,.
 $$

\subsection{Probability-density bifurcation}

Incidentally, all of the four minima of our $k=7$ potential $V(x)$
happen to be equally deep at $\alpha=\beta=\gamma=1$. This is a
random coincidence. Due to Eqs.~(\ref{35}) and (\ref{37}), such a
feature of the model remains valid along the whole one-parametric
line of $\beta=\gamma= \sigma\alpha$. The proof of this observation
follows from the comparison of the exact formula
 \be
 V(\pm \alpha)=-\left( {1+12\,\sigma}^{2}+36\,{\sigma}^{4}\right)\,
 \alpha^{8}\,
 \ee
for the depth of the potential at its inner minima with formula
 \be
 V(\pm R)=- {\alpha}^{8}  (1+6\sigma^2)^{2} \,
 \ee
for the depths of the outer valleys.

At the asymptotically large parameters (i.e., say, outside of a
large ball of radius $\varrho$), the depth $ V(\alpha)= {\cal
O}(\varrho^8)$ of the valleys (i.e., the large and negative
leading-order contribution to the energy) is by five orders of
magnitude larger than the distances $\sim \omega = {\cal
O}(\varrho^3)$ between the separate energy levels. Thus, the quantum
effects remain comparatively small in the asymptotic domain of large
$\varrho$. The size of the depth $V(\alpha)$ remains a decisive
criterion for the dominance or suppression of the locally supported
wave functions $\psi_n(x)$. This observation strictly parallels the
analogous feature of the preceding $k=5$ model.

Once we break the balance between $\beta= \sigma\alpha$ and $\gamma=
\sigma\alpha$, the fragile four-centered balance in probability
density will also break down. At $\beta\neq \gamma$ this density may
be expected to concentrate near the inner centers of $x \approx
x^{\rm (equilibrium)}_1= \pm \alpha$, or near the two more remote
outer centers of $x \approx x^{\rm (equilibrium)}_2= \pm R = \pm
\alpha\,\sqrt{1+6\sigma^2}$. The process can be interpreted as a
quantum analogue of classical bifurcation, and the higher-order
corrections in $1/\alpha$ can enter the game at the smaller radii
$\varrho$.

One of the characteristic consequences of the spatial symmetry of
our $k=7$ model is the delocalized, two-centered nature of the inner
and outer equilibria. Whenever $\beta \neq \gamma$, these equilibria
remain stable with respect to small perturbations. From this
perspective the genuine quantum catastrophe (i.e., the inner-well --
outer-outer jump) will be encountered during the passage of the
quantum system in question through its $\beta=\gamma$ interface.

It is instructive to notice that in the leading order approximation
the discreteness (i.e., the quantum-theory nature) of the
equilibrium does not play any significant role. The subdominant
corrections must enter the game in the sub-asymptotic parametric
domain. This leads to amended formulae for energies, clearly
separated into the inner-wells-supported sub-spectrum
 \be
 E_n^{(inner-wells)}=V(\alpha)+(2n+1) \omega+ \ldots\,,
 \ \ \ \ \ n=0,1, \ldots
 \label{iwe}
 \ee
and its outer-wells complement
 \be
 E_n^{(outer-wells)}=V(R)+(2n+1) \Omega+ \ldots\,,
 \ \ \ \ \ n=0,1, \ldots\,.
 \label{owe}
 \ee
In each of these subsets, the weight of the subdominant corrections
could be enhanced by the transition to the systems with large
$\Lambda=\hbar/\sqrt{2\mu}$ (i.e., with small mass $\mu$), but for
the time being let us keep the latter parameter fixed at its
conventional value of $\Lambda=1$. Then, an inclusion of subdominant
corrections still becomes obligatory, say, at the non-asymptotic
values of the cut-off radius $\varrho$. Even though the approximate
degeneracy of the parity doublets would also become less and less
pronounced due to such a decrease of $\varrho$, it still remains
less essential. Nevertheless, one has to proceed with due care. For
example, whenever we insist on the reduction
$\beta=\gamma=\sigma\alpha$, we get the $n-$th-state energy
difference
 \be
 \Delta_n=E_n^{(outer-wells)}-E_n^{(inner-wells)}=
 (\Omega-\omega)(2n+1)=
 12(2n+1)\,\alpha^3\,\sigma^2
 \left(
 \sqrt{1+6\sigma^2}-1
 \right)
 \label{pjunta}
 \ee
which is a strictly positive quantity which cannot vanish. The
perceivably narrower outer valleys are always pushing the outer
spectrum up. For this reason the catastrophe cannot be reached
unless we leave the two-dimensional $\beta=\gamma$ surface.

\begin{figure}[h]                    
\begin{center}                         
\epsfig{file=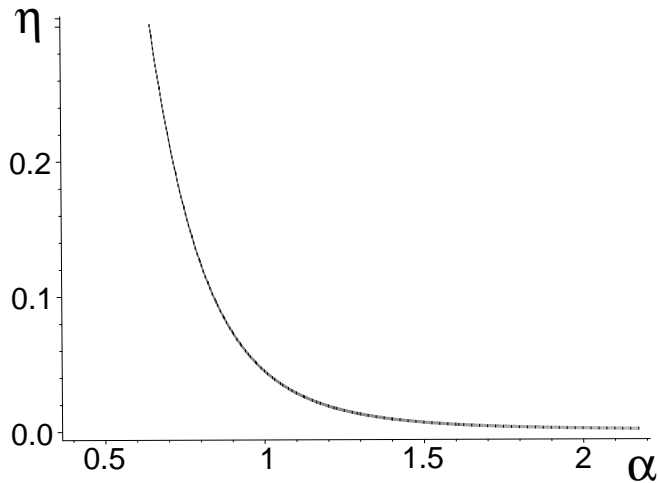,angle=270,width=0.56\textwidth}
\end{center}    
\vspace{2mm} \caption{The quantum-catastrophe interface $\eta^{(
critical)}=\eta^{( critical)}(\alpha)$ calculated at $\beta=\alpha$
and at the slightly enhanced $\gamma=(1+\eta)\alpha$ using
approximations (\ref{iwe}) and (\ref{owe}).
 \label{staja}
 }
\end{figure}

In a characteristic numerical experiment let us set, therefore,
$\beta=\alpha$ and $\gamma=(1+\eta)\alpha$ with a small and positive
$\eta$. This enables us to treat the definition of the quantum
relocalization catastrophe as the specification of the enhancement
parameter $\eta^{( critical)}$, studied as the function of the
second free parameter $\alpha$. In spite of the rather complicated
form of the explicit expression for the energy difference
$\Delta(\alpha,\eta)$ in the given approximation, the shape of the
resulting function $\eta^{( critical)}(\alpha)$ remains surprisingly
smooth though still consistent with our expectations (cf.
Fig.~\ref{staja}). The critical value of $\eta$ keeps to be very
small even at the not too large large $\alpha$. With the decrease of
the latter value below cca $\alpha \approx 1.5$ the critical curve
$\eta$ starts growing more quickly.


\section{Discussion\label{k6}}


\subsection{Constructive and spectral-theory context\label{krkk6}}

Our numerical experiments performed at $k=7$ or higher lead us to a
rather sceptical attitude towards the reliability of the mere
second-order perturbation-theory analysis. Naturally, such an
approximation proved perfectly sufficient for our present purposes
and analysis of systems living in the extreme dynamical regimes.
Nevertheless, in a more general setting such a technique can hardly
be perceived as sufficiently reliable in the {\em whole\,} space of
parameters, especially when one decides to move far from the initial
asymptotic domain.

For the more universal and reliable quantitative predictions the
higher-order corrections would have to be included. In the related
detailed studies one could also try to search for help in the
extensive existing literature. Besides an eligible turn of attention
to the brute-force numerical calculations (which may be sampled,
e.g., by the old but still relevant case study \cite{oldqc}) and
besides recalling the early rigorous uses of the harmonic-oscillator
approximations accompanied by an estimate of errors \cite{Olver}, an
important source of inspiration of a way towards amendments should
be sought in the incessant progress in our understanding of the
semiclassical and Stokes-geometry methods \cite{cheb1}.

In parallel, a remarkable methodical encouragement emerges also from
the correspondence between the bound-state and scattering problems,
especially due to an intimate relationship between the necessary
matching formulae in the systems with and without scattering
\cite{hauge}. The analogies become particularly striking when one
considers a particle passing through several separate barriers
\cite{cheb2}. In this context, a technical inspiration might also
result from certain mathematically analogous situations occurring in
classical optics \cite{han1,stoplight} or in classical
electrodynamics \cite{stov1}.


\subsection{A remark on the use of classical catastrophes
in quantum world\label{ckkk6}}

Knowing that the perturbation approximations as well as their
semiclassical parallels might fail even at the deceptively small
values of the first-order corrections, the last resort still remains
to lie in the determination of the relocalization catastrophes via a
suitable numerical, brute-force integration of the underlying
Schr\"{o}dinger equation with a carefully controlled precision.
Still, our understanding of these results may rely upon the
classical classification schemes (cf., e.g., \cite{revcat}).

In this setting we have to re-emphasize that the notion of
catastrophe was really introduced as a purely classical, non-quantum
concept. Indeed, the popular Thom's catastrophe theory characterizes
the static equilibria of dynamical systems as minima of polynomials.
The theory covers, systematically, the scenarios in which one
encounters an abrupt, thorough change of the system caused by a
small change of parameters. In our present paper we studied, in a
concise characterization, just one of the possibilities of an
upgrade of such a notion and of its transfer in quantum world.

Our study was inspired by the Arnold's \cite{Arnold} list of the
classical catastrophe-related polynomials $V_{(k)}(x) = x^{k+1} +
\ldots$. We considered just the one-dimensional confined motion of a
single massive particle in such a potential, and we replaced its
classical dynamics laws simply by the quantum ones. In this
framework we had to recall the well known facts that in quantum
mechanics, in contrast to classical mechanics, the
polynomial-potential barriers are never impenetrable, and that the
ground-state energy level can only coincide with the (absolute)
minimum of the potential in the semi-classical {\it alias\,}
infinitely-heavy-particle limit.


\subsection{Relocalization catastrophes
\label{ck6}}

On the latter background we emphasized that the phenomenon of
tunneling leads to the fundamental and irreparable instability of
any quantum system when the dominant power exponent $k+1$ is odd. In
such cases the instability is caused by the asymptotically
antisymmetric shape of potentials $V_{(2N)}(x) = x^{2N+1} + c_1
x^{2N-1}+ \ldots$. Thus, we could only pay attention to the Arnold's
potentials with even exponents $k+1=2N+2$. For the sake of formal
simplicity the latter family was further restricted here to the mere
spatially symmetric potentials $V_{(2N+1)}(x)=x^{2N+2}+ c_1
x^{2N}+\ldots$ with $c_2=c_4=\ldots =0$. At the same time, the
practical phenomenological appeal of these potentials was enhanced
by a sign-changing convention $c_1<0$, $c_3>0$, $c_5<0$, \ldots and
by a suitable reparametrization of these couplings.

The latter constraint enabled us to keep the number of the barriers
in $V(x)$ maximal, equal to $N$. This means that the resulting
models exhibited a remarkable geometrical shape-description economy.
This merit was also complemented by a mathematical
round-off-suppression feature. A rather practical user-friendliness
of the formalism was achieved via the requirement of having all of
the relevant polynomial formulae represented in non-fractional
arithmetics, i.e., using just integer coefficients.

As long as the related combinatorial analysis proved rather
technical, its detailed description beyond $N=4$ was moved from the
bulk text to the Appendix. Anyhow, the results of our
straightforward systematic computer-assisted diophantine analysis
proved easily feasible up to the Arnold's polynomials $V(x)$ of
degree eighteen.


\subsection{More general concepts of quantum catastrophes
\label{cqk6}}

In a quantum-theory context of Schr\"{o}dinger equation we treated
the mass-parameter ${{\Lambda}}={\hbar}/\sqrt{2\mu}$ as a variable
quantity, in principle at least. This facilitated a clarification of
the classical-quantum physics parallel. Indeed, first of all, one
can formally get the classical system by merely letting the small
parameter ${{\Lambda}}$ decrease to zero. Although the opposite,
classical-to-quantum correspondence is less obvious and may be
ambiguous, we emphasized its relevance.

Elementary observations of the latter type were shown to imply,
e.g., that due to the emergence of the tunneling, the classical
equilibria and bifurcation catastrophes called ``fold'' {\em
cannot\,} have any stable quantum analogue. Similarly, even in the
stable quantum analogue of the next Thom's classical catastrophe
called ``cusp'', the tunneling will smear out the bifurcation
phenomenon completely. For these reasons, one has to be rather
careful with the terminology. In the current literature, indeed, one
finds several non-equivalent concepts of the catastrophic dynamics
ranging from the so called ``orthogonality catastrophe''
\cite{orthocat} (with its origin dating back to the well known
Anderson's orthogonality theorem \cite{Anders}) up to the various
{\it ad hoc\,}} forms of the descriptions of quantum phase
transitions \cite{catopen,nonEP,catanucl}.

A purely pragmatic resolution of the latter ambiguities has been
found here in the ultimate restriction of our attention to the
specific, ``local deep well'' dynamical regime. For the purposes of
building the theory this gave us the two decisive advantages.
Firstly, the restriction enabled us to simplify the mathematics, in
essence, by the quick associated decrease of the size of the
corrections to the dominant and exactly solvable harmonic-oscillator
leading-order approximations. Secondly, the physics proved clarified
precisely due our initially purely technical assumption of the
spatial symmetry of $V(x)=V(-x)$. Thus, we could turn attention to
the almost degenerate doublets of levels distinguished, up to
negligible errors, just by their parity.

In this sense, the above-mentioned phenomena of tunneling leading to
certain ``no-go'' statements about the absence of bifurcations has
been shown to remain restricted just to the most elementary
cusp-related $V_{(2M+1)}(x)$ with $N=1$ (i.e., with the single
barrier). We showed that in multi-well potentials there emerges the
possibility of having the quantum bifurcation phenomenon involving
the different-parity energy doublets rather than the single
non-degenerate states themselves.


\subsection{Summary\label{kkk6}}

In our present approach, quantum catastrophes are perceived as
phenomena which occur in a one-dimensional single-particle quantum
system with the dynamics controlled by a suitable local potential
$V(x)$. Then, the catastrophe itself is characterized by an abrupt
relocalization of the probability density. Under these assumptions
we showed that a measurable relocalization is only possible at
parameters $N \geq 2$.

For the sake of simplicity of our considerations we only considered
the Arnold's menu of potentials. As a consequence, out of the
popular Thom's list, only the ``butterfly'' option survived, with
$V_{(5)}(x)$. Under this restriction we have shown, constructively,
that the practical feasibility features of the latter $N=2$
benchmark model remain fully preserved at $N=3$, etc. Several
explicit second-order-precision examples of the relocalization were
also presented for illustration.

Summarizing, one of the key messages delivered by the present paper
is that quite a few models with $N \geq 4$ are still comparably
easily tractable. Thus, in principle, they seem to admit a
non-numerical treatment leading to a rather universal and
unexpectedly user-friendly classification scheme. For the purposes
of applications it would be only necessary to replace the standard
hand-made, pencil and paper style of working with polynomials by the
currently commercially available computer-assisted symbolic
manipulation techniques.

\newpage

\section*{Appendix A: Parametrizations of $V(x)$ up to $k = 17$}

%
%

\begin{figure}[h]                    
\begin{center}                         
\epsfig{file=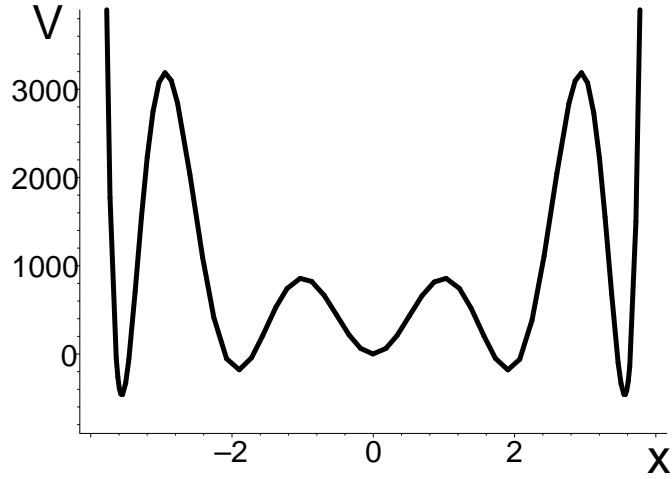,angle=270,width=0.56\textwidth}
\end{center}    
\vspace{2mm} \caption{Five-well potential (\ref{kaje9}) with
$\alpha=1$, $\beta=\sqrt{4/6}$, $\gamma=\sqrt{5/6}$, and
$\delta=\sqrt{2/6}$ in (\ref{45}).
 \label{wja}
 }
\end{figure}
%

\subsection*{A.1. $V(x)$ with four barriers ($k=9$)}

The Arnold's $k=9$ potential in its spatially symmetrized
four-parametric form
 \be
 V(x) = x^{10}-5a^2x^8+10b^2x^6-10c^2x^4+5d^2x^2
 \label{kaje9}
 \ee
is, thanks to the auxiliary binomial coefficients, easily
differentiated,
 \be
 V'(x) = 10\,
 \left (\xi^{4}-4a^2\xi^3+6b^2\xi^2-4c^2\xi+d^2\right ) x=
  10 \,x\cdot C^{(4)}(\xi)\,,
  \ \ \ \ \xi=x^2\,.
 \label{[25b]}
 \ee
In a way which parallels Eqs.~(\ref{elik}) and (\ref{gelik}) we
found the following optimal ansatz
  \be
  C^{(4)}(\xi)=
 \left( \xi-{\alpha}^{2} \right)  \left(
 \xi-{\alpha}^{2}-4\,{\beta}^{2} \right)  \left(
 \xi-{\alpha}^{2}-4\,{\beta}^{2}-6\,{\gamma}^{2}
 \right)  \left( \xi-{\alpha}^{2}-4\,{\beta}^{2}-6\,{\gamma}^{2}-12\,
 \delta^2 \right)\,.
 \label{26}
 \ee
Its specification was based on the requirement of a maximal
simplification of the formulae for the extremes of $V(x)$. Product
(\ref{26}) may be expanded into polynomial and compared with
Eq.~(\ref{[25b]}). As long as we used computer-assisted symbolic
manipulations, we easily obtained a lengthier but still printable
polynomial in $\xi$,
 \ben
  C^{(4)}=
  {\xi}^{4}+ \left( -12\,{\delta}^{2}-12\,{\beta}^{2}-12\,{\gamma}^{2}-4
 \,{\alpha}^{2} \right) {\xi}^{3}+
 \een
 \ben
 +\left(
 36\,{\gamma}^{4}+96\,{\beta}^
 {2}{\delta}^{2}+96\,{\beta}^{2}{\gamma}^{2}+36\,{\alpha}^{2}{\beta}^{2
 }+36\,{\alpha}^{2}{\delta}^{2}+48\,{\beta}^{4}+72\,{\gamma}^{2}{\delta
 }^{2}+6\,{\alpha}^{4}+36\,{\alpha}^{2}{\gamma}^{2} \right)
 {\xi}^{2}+
  \een
  \ben
  +
 \left( -288\,{\beta}^{2}{\gamma}^{2}{\delta}^{2}-144\,{\beta}^{2}{
 \gamma}^{4}-144\,{\alpha}^{2}{\gamma}^{2}{\delta}^{2}-4\,{\alpha}^{6}-
 72\,{\alpha}^{2}{\gamma}^{4}-192\,{\beta}^{4}{\gamma}^{2}-96\,{\alpha}
 ^{2}{\beta}^{4}-\right .
 \een
 \ben
 \left . -36\,{\alpha}^{4}{\beta}^{2}
 -36\,{\alpha}^{4}{\delta}^{2}
 -192\,{\alpha}^{2}{\beta}^{2}{\gamma}^{2}-64\,{\beta}^{6}-36\,{
 \alpha}^{4}{\gamma}^{2}-192\,{\alpha}^{2}{\beta}^{2}{\delta}^{2}-192\,
 {\beta}^{4}{\delta}^{2} \right)
 \xi+
 \een
 \ben
 + 48\,{\alpha}^{4}{\beta}^{4}+36\,{
 \alpha}^{4}{\gamma}^{4}+12\,{\alpha}^{6}{\delta}^{2}+12\,{\alpha}^{6}{
 \gamma}^{2}+144\,{\alpha}^{2}{\beta}^{2}{\gamma}^{4}+64\,{\alpha}^{2}{
 \beta}^{6}+72\,{\alpha}^{4}{\gamma}^{2}{\delta}^{2}+
 \een
 \be
 +
 192\,{\alpha}^{2}{
 \beta}^{4}{\gamma}^{2}+96\,{\alpha}^{4}{\beta}^{2}{\gamma}^{2}+192\,{
 \alpha}^{2}{\beta}^{4}{\delta}^{2}+12\,{\alpha}^{6}{\beta}^{2}+96\,{
 \alpha}^{4}{\beta}^{2}{\delta}^{2}+288\,{\alpha}^{2}{\beta}^{2}{\gamma
 }^{2}{\delta}^{2}+{\alpha}^{8}\,.
 \label{rela}
 \ee
This expression must coincide with polynomial (\ref{[25b]}). The
comparison defines the original couplings in (\ref{kaje9}) in terms
of the new parameters,
 \ben
 a^2={ \alpha}^{2}+
 3\,{\beta}^{2}+3\,{\gamma}^{2}+3\,\delta^2\,,
 \een
 \ben
 b^2=
 6\,{\alpha}^{2}{\delta}^{2}+16\,{\beta}^{2}{{\gamma}}^{2}+{\alpha}^{4}+12\,{{\gamma}
 }^{2}{\delta}^{2}+6\,{{\gamma}}^{4}+6\,{\alpha}^{2}{{\gamma}}^{2}+16\,{\beta}^{2}{
 \delta}^{2}+8\,{\beta}^{4}+6\,{\alpha}^{2}{\beta}^{2}\,,
 \een
 \ben
 c^2=9\,{\alpha}^{4}{{\gamma}}^{2}+16\,{\beta}^{6}+9\,{\alpha}^{4}{\beta}^{2}+36
\,{\alpha}^{2}{{\gamma}}^{2}{\delta}^{2}+24\,{\alpha}^{2}{\beta}^{4}+18\,{
\alpha}^{2}{{\gamma}}^{4}+48\,{\alpha}^{2}{\beta}^{2}{\delta}^{2}+
\een
 \ben
 +36\,{\beta}
^{2}{{\gamma}}^{4}+{\alpha}^{6}+48\,{\beta}^{4}{\delta}^{2}+9\,{\alpha}^{4}{
 \delta}^{2}+72\,{\beta}^{2}{{\gamma}}^{2}{\delta}^{2}+48\,{\beta}^{4}{{\gamma}}^{2}+
 48\,{\alpha}^{2}{\beta}^{2}{{\gamma}}^{2}\,,
 \een
 \ben
 d^2=144\,{\alpha}^{2}{\beta}^{2}{{\gamma}}^{4}+12\,{\alpha}^{6}{{\gamma}}^{2}+12\,{
 \alpha}^{6}{\beta}^{2}+48\,{\alpha}^{4}{\beta}^{4}+12\,{\alpha}^{6}{
 \delta}^{2}+96\,{\alpha}^{4}{\beta}^{2}{{\gamma}}^{2}+
 \een
 \ben
 +64\,{\alpha}^{2}{\beta}
 ^{6}+36\,{\alpha}^{4}{{\gamma}}^{4}+192\,{\alpha}^{2}{\beta}^{4}{{\gamma}}^{2}+192\,
 {\alpha}^{2}{\beta}^{4}{\delta}^{2}+{\alpha}^{8}+72\,{\alpha}^{4}{{\gamma}}^{
 2}{\delta}^{2}+
 \een
  \be
  +288\,{\alpha}^{2}{\beta}^{2}{{\gamma}}^{2}{\delta}^{2}+96\,{
 \alpha}^{4}{\beta}^{2}{\delta}^{2}\,.
 \label{45}
 \ee
All of these expressions are polynomials with integer coefficients.
In fact, the complete suppression of the fractional coefficients was
precisely the purpose of the {\it ad hoc\,} scaling in (\ref{26}).
At $N=3$ the demonstration of the uniqueness of such a scaling was
shown in paragraph \ref{uhu} at $N=3$. It is also unique at $N=4$ --
the proof can be performed, quickly, via formal replacements $4 \to
P$, $6 \to T$ and $12 \to V$ in (\ref{26}). In the resulting
generalized expansion (\ref{rela}), the divisibility of the
coefficient at $\xi^3$ (by four) confirms the minimality of $P=4$.
Subsequently, the divisibility constraint at $\xi^2$ (by six) proves
the minimality of $T=6$ as well as of $V=12$. The last divisibility
condition at $\xi$ (by four) appears then already satisfied ``for
free'', without imposing any additional constraints.

The practical benefits provided by this computer-assisted
diophantine analysis are obvious. Several deeper combinatorial
aspects of this result remain still unexplained and challenging. For
example, the necessity of having the weight ``12'' at the outermost
shift-parameter $\delta^2$ disproved a tentative
``binomial-coefficients'' extrapolation hypothesis inspired by
Eq.~(\ref{elik}).

An $N=4$ anomaly emerges also in the shape of the potential because
after the ``trivial'' choice of $\alpha=\beta=\gamma=\delta=1$ the
$k=9$ graph of $V(x)$ appears dominated by the pronounced outermost
pair of the very deep absolute minima. Again, there is no analogy
with the $N=3$ case of Fig.~\ref{fifi}. This loss of analogy is
slightly unfortunate because for the purposes of a localization of
the quantum catastrophic dynamical regime, the unwanted dominance of
the outer minima must be suppressed, e.g., via an {\it ad hoc\,}
decrease of $\delta^2$. A sample of the results of such a tentative
suppression is given, in Fig.~\ref{wja}, for potential
 $$
 V(x)=
 {x}^{10}-{\frac {65}{2}}\,{x}^{8}+355\,{x}^{6}-{\frac {39860}{27}}\,{x
 }^{4}+{\frac {54340}{27}}\,{x}^{2}
  $$
with $\beta^2={2/3}$ $\gamma^2={5/6}$ and $\delta^2={1/3}$.



\begin{figure}[h]                    
\begin{center}                         
\epsfig{file=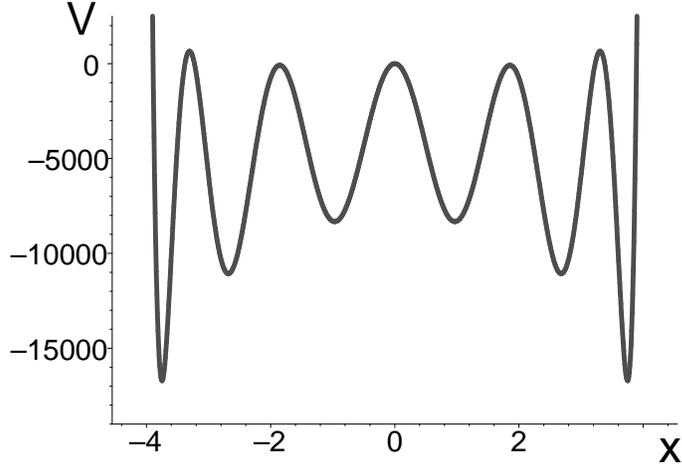,angle=270,width=0.56\textwidth}
\end{center}    
\vspace{2mm} \caption{Six-well potential (\ref{kaje11}) with
$4\alpha=\sqrt{15}$, $4\beta=\sqrt{8}$, $4\gamma=4\delta=\sqrt{6}$
and $4\epsilon=\sqrt{5}$.
 \label{vojel}
 }
\end{figure}
%


\subsection*{A.2. $V(x)$ with five barriers ($k=11$)}

The problem of the localization of the maxima and minima in the
$k=11$ potential
 \be
 V(x) = x^{12}-6a^2x^{10}+15b^2x^8-20c^2x^6+15d^2x^4-6f^2x^2
 \label{kaje11}
 \ee
leads, through differentiation
 \be
 V'(x) = 12\,
 \left (\xi^{5}-5a^2\xi^4+10b^2\xi^3-10c^2\xi^2+5d^2\xi-f^2\right ) x=
  12 \,x\cdot C^{(5)}(\xi)
 \label{25}
 \ee
to the factorization ansatz
  \ben
 C^{(5)}(\xi)=
 \left( \xi-{\alpha}^{2} \right)  \left( \xi-{\alpha}^{2}-P{\beta}^{2} \right)
  \left( \xi-{\alpha}^{2}-P{\beta}^{2}-Q{g}^{2} \right)
 \left( \xi-{\alpha}^{2}-P{\beta}^{2}-Q{g}^{2}-R{\delta}^{2} \right)
 \times
 \een
 \be
 \times
 \left(
 \xi-{\alpha}^{2}-P{\beta}^{2}-Q{g}^{2}-R{\delta}^{2}-S\epsilon^2
 \right)\,.
 \ee
In its optimal realization we have to choose $2P=Q=R=S=10$,
 \ben
 C^{(5)}(\xi)=
  \left( \xi-{\alpha}^{2} \right)  \left(
 \xi-{\alpha}^{2}-5\,{\beta}^{2} \right)
 \left(
 \xi-{\alpha}^{2}-5\,{\beta}^{2}-10\,{{\gamma}}^{2}
 \right) \times
 \een
 \be
 \times \left( \xi-{\alpha}^{2}-5\,{\beta}^{2}-10\,{{\gamma}}^{2}-10\,{
 \delta}^{2} \right)  \left(
 \xi-{\alpha}^{2}-5\,{\beta}^{2}-10\,{{\gamma}}^
 {2}-10\,{\delta}^{2}-10\,{\epsilon}^{2} \right)\,.
 \label{beben}
 \ee
The technique of the determination of the optimal scaling
coefficients remains the same as above.

Along the same lines as above, one also extracts the explicit form
of the general six-well analogue of Eq.~(\ref{45}) deduced from
Eq.~(\ref{beben}) and yielding
 \be
 a^2=
 {\alpha}^{2}+4\,{\beta}^{2}+6\,{{\gamma}}^
 {2}+4\,{\delta}^{2}+2\,{\epsilon}^{2}\,
 \ee
etc. These explicit formulae for the reparametrized coupling
constants still remain sufficiently compact to help us to test, vary
or fine-tune the shape of the potential in its dependence on
parameters $\alpha$ - $\epsilon$. A sample of this shape is
presented in Fig.~\ref{vojel}.

\subsection*{A.3. $V(x)$ with six barriers ($k=13$) }

Once we decide to avoid the presence of fractional coefficients in
the definitions of couplings at $k=13$, we may apply the same recipe
as above and to factorize, in an optimal manner,
 \ben
 C^{(6)}(\xi)=
 \left( \xi-{\alpha}^{2} \right)  \left( \xi-{\alpha}^{2}-6\,{\beta}^{2} \right)
  \left( \xi-{\alpha}^{2}-6\,{\beta}^{2}-15\,{{\gamma}}^{2}
 \right)  \left( \xi-{\alpha}^{2}-6\,{\beta}^{2}-15\,{{\gamma}}^{2}-20\,{
 \delta}^{2} \right) \times
 \een
 \be
 \times
  \left(
 \xi-{\alpha}^{2}-6\,{\beta}^{2}-15\,{{\gamma}}^{2
 }-20\,{\delta}^{2}-30\,{\epsilon}^{2} \right)  \left(
 \xi-{\alpha}^{2}
 -6\,{\beta}^{2}-15\,{{\gamma}}^{2}-20\,{\delta}^{2}-30\,{\epsilon}^{2}-60\,{
 \zeta}^{2} \right)\,.
 \ee
Again, the patient diophantine analysis gave us the result which
couldn't have been guessed in advance. It leads to the couplings
expressed as polynomials with integer coefficients, starting from
 \be
 a^2=
 {\alpha}^{2}+5\,{\beta}^{2}+10\,{{\gamma}}^
 {2}+10\,{\delta}^{2}+10\,{\epsilon}^{2}+10\,{\zeta}^{2}\,.
 \ee
The complete list of formulae would be, unfortunately, too long for
a display in print.




\subsection*{A.4. $V(x)$ with seven and eight barriers ($k=15$ and $17$) }

Our preceding diophantine analysis revealed an irregularity in the
$N-$ or $k-$dependence of the optimal ansatzs for the positions of
the minima and of the maxima of the present subfamily of Arnold's
potentials $V(x)$. Several extrapolation hypotheses were also found
incorrect when we further proceeded to $N=7$, i.e., to the
factorized polynomial
 \ben
  C^{(7)}(\xi)=
\left( \xi-{\alpha}^{2} \right)  \left( \xi-{\alpha}^{2}-7\,{\beta}^{2} \right)
 \left( \xi-{\alpha}^{2}-7\,{\beta}^{2}-21\,{{\gamma}}^{2}
 \right)  \left( \xi-{\alpha}^{2}-7\,{\beta}^{2}-21\,{{\gamma}}^{2}-105\,{
\delta}^{2} \right)
\times
\een
 \ben
 \times \left( \xi-{\alpha}^{2}-7\,{\beta}^{2}-21\,{{\gamma}}^{2
}-105\,{\delta}^{2}-35\,{\epsilon}^{2} \right)  \left( \xi-{\alpha}^{2
}-7\,{\beta}^{2}-21\,{{\gamma}}^{2}-105\,{\delta}^{2}-35\,{\epsilon}^{2}-105
\,{\zeta}^{2} \right)
\times
\een
\ben
\times
 \left( \xi-{\alpha}^{2}-7\,{\beta}^{2}-21\,{{\gamma}}^
{2}-105\,{\delta}^{2}-35\,{\epsilon}^{2}-105\,{\zeta}^{2}-105\,{\eta}^
{2} \right)\,.
 \een
Its computer-assisted expansion yield again the formulae for the
couplings, say,
 \be
 a^2=
 {\alpha}^{2}+6\,{\beta}^{2}+15\,{{\gamma}}^
 {2}+60\,{\delta}^{2}+15\,{\epsilon}^{2}+30\,{\zeta}^{2}+15\,{\eta}^{2}\,
 \ee
etc. Similarly, our final, $N=8$ combinatorial auxiliary result
reads
%
%
%
 \ben
  C^{(8)}(\xi)=
 \left( \xi-{\alpha}^{2} \right)  \left( \xi-{\alpha}^{2}-8\,{\beta}^{
2} \right) \left( \xi-{\alpha}^{2}-8\,{\beta}^{2}-28\,{{\gamma}}^{2}
 \right)
 \times \ldots
 \een
 \be
 \ldots
 \times \left( \xi-{\alpha}^{2}-8\,{\beta}^{2}-28\,{{\gamma}}^{2}-56\,{
\delta}^{2}-70\,{\epsilon}^{2}-280\,{\zeta}^{2}-140\,{\eta}^{2}-280\,
\theta \right)
 \ee
and implies, that
 \be
 a^2=
 {\alpha}^{2}+7\,{\beta}^{2}+
 21\,{{\gamma}}^
 {2}+35\,{\delta}^{2}+35\,{\epsilon}^{2}+105\,{\zeta}^{2}
 +35\,{\eta}^{2}+35\,{\theta}^{2}\,
 \ee
etc. None of these formulae looks amenable to an easy extrapolation
in $N$. At the same time we found that whenever needed, their direct
computer-assisted constructions still remains also very quick, at
the next few integers $N>8$ at least.

\end{document}